# Graphene hot-electron light bulb: incandescence from hBN-encapsulated graphene in air

Seok-Kyun Son[1,4], Makars Šiškins[2,4], Ciaran Mullan[2,4], Jun Yin[2], Vasyl G. Kravets[2], Aleksey Kozikov[1], Servet Ozdemir[2], Manal Alhazmi[2], Matthew Holwill[1], Kenji Watanabe[3], Takashi Taniguchi[3], Davit Ghazaryan[1], Kostya S. Novoselov[1,2], Vladimir I. Fal'ko[1,2], Artem Mishchenko[1,2,5]

[1]*National Graphene Institute, University of Manchester, Manchester M13 9PL, UK*

[2]*School of Physics and Astronomy, University of Manchester, Manchester M13 9PL, UK*

[3]*National Institute for Materials Science, 1-1 Namiki, Tsukuba, 305-0044, Japan*

[4] These authors contributed equally

[5] e-mail: artem.mishchenko@gmail.com

**Abstract:** The excellent electronic and mechanical properties of graphene allow it to sustain very large currents, enabling its incandescence through Joule heating in suspended devices. Although interesting scientifically and promising technologically, this process is unattainable in ambient environment, because graphene quickly oxidises at high temperatures. Here, we take the performance of graphene-based incandescent devices to the next level by encapsulating graphene with hexagonal boron nitride (hBN). Remarkably, we found that the hBN encapsulation provides an excellent protection for hot graphene filaments even at temperatures well above 2000 K. Unrivalled oxidation resistance of hBN combined with atomically clean graphene/hBN interface allows for a stable light emission from our devices in atmosphere for many hours of continuous operation. Furthermore, when confined in a simple photonic cavity, the thermal emission spectrum is modified by a cavity mode, shifting the emission to the visible range spectrum. We believe our results demonstrate that hBN/graphene heterostructures can be used to conveniently explore the technologically important high-temperature regime and to pave the way for future optoelectronic applications of graphene-based systems.

## 1. Introduction

The concept of van der Waals heterostructures has led to a new technology of layer-by-layer engineering of two-dimensional (2D) materials with atomic precision [1, 2]. Out of the abundance of 2D materials, the two – graphene and hexagonal boron nitride (hBN), still remain the most unique couple, because of their outstanding synergetic properties. The hBN plays a vital role in ultra-high quality graphene devices [3, 4], and as an enabling material for graphene superlattices [5]. Besides, it is also known to protect sensitive 2D materials from environment via encapsulation [6]. Graphene, among its many superlatives, has the highest breakdown current density $j_{max} \sim 10^{12}$ A/cm² (assuming graphene thickness 3.4 Å), as measured in transmission experiments of accelerated (180 keV) xenon ions [7]. In electronic transport devices, the breakdown current densities of graphene are more modest, reaching ~ $5 \cdot 10^8$ A/cm$^2$ in a vacuum (and much lower values in air) [8].

The high current-carrying capacity of graphene has made it a simple source of thermal infrared emission [9, 10]. Controlled by a bias voltage, a maximal emission temperature is limited to ~ 700 K due to heat dissipation to the metallic contacts and to the substrate [8, 10]. The latter mechanism dominates for graphene devices longer than a few microns. In the case of hBN encapsulated graphene devices, heat can also be efficiently dissipated through radiative heat transfer to hBN via its hyperbolic phonon polaritons [11]. Heat transfer is substantially reduced for suspended graphene devices, which leads to much higher emission temperatures approaching 3000 K, and to the emission in the visible spectrum (incandescence) when biased by electrical pulses under high vacuum [12]. But the stability of these devices in air is severely limited due to the oxidation of graphene [8, 12].

Metallic single-walled carbon nanotubes were also used as thermal light emitters, although, small saturation currents (< 20 µA for a 3-4 nm diameter nanotube) and rapid oxidation in the air limit their use [13]. Interestingly, the electronic transitions between van Hove singularities led to a strong modification of the emitted spectra [13], which can be tuned even further by employing one-dimensional photonic crystal cavities [14]. In this work, the encapsulation of graphene (Gr) with hBN allowed us to demonstrate robust incandescence of graphene devices in ambient conditions under continuous DC bias over extended periods of time. The hBN encapsulation provides excellent protection for graphene even at temperatures well above 2000 K, which we attribute to an unparalleled oxidation resistance of hBN [15-18]. Our study also reveals that dielectric cavity can support surface-guided waves at visible wavelength and tune the spectra of Planckian radiation.

## 2. Light emission from hBN-encapsulated graphene

Figure 1a shows one of our hBN/Gr/hBN heterostructures fabricated on quartz substrate, using conventional dry peel technology followed by a standard e-beam lithography, dry etching and metallisation, as described elsewhere [4, 19]. We designed our devices such that the distance between the contacts is long enough ($L$ ~ 10 µm) to avoid any effects from the contacts while keeping the contact resistance low by increasing the metal-to-graphene contact length. To characterise the light emission we pass a large current $I$ (current density, $j = I/W$, reaching 600 A/m) through a narrow ($W$ ~ 1 µm is the channel width) graphene channel by applying a bias voltage $V_b$ (estimated electric field $E = V_b/L$ ~ 1.5 V/µm, where $L$ is the channel length). The supplied electric energy is transformed into Joule heat and dissipated in graphene, encapsulating hBN, quartz substrate, and metallic contacts, leaving a small fraction to radiate into free space. Figure 1b shows the emission spectra recorded using Renishaw Raman spectrometer equipped with Si-based CCD detector (diffraction grating 1200/mm; edge filter and beam splitter were removed during measurements). The obtained spectra are mostly featureless and can be described using Planck's law, modified by energy independent emissivity $\epsilon$ (grey-body radiation) [9]:

$$I_{grey-body}(h\nu, T) = \epsilon \frac{2(h\nu)^3}{h^2c^2} \frac{1}{\exp\left(\frac{h\nu}{k_BT}\right)-1} \quad , \tag{1}$$

where $h\nu$ is the photon energy, $T$ is the temperature, $c$ is the speed of light, $k_B$ and $h$ are Boltzmann's and Planck's constants.

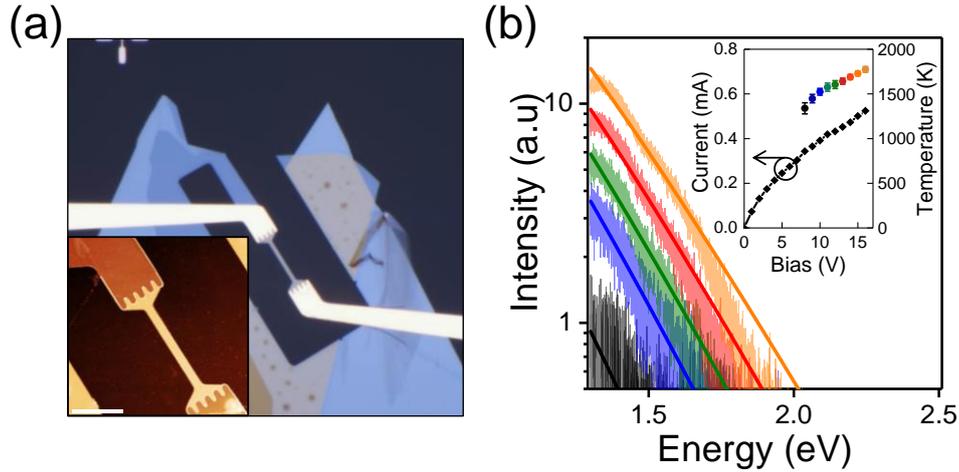

**Figure 1.** Thermal radiation of Joule-heated hBN/Gr/hBN device on quartz substrate, measured in ambient conditions. (a) Optical micrograph of hBN-encapsulated graphene filament. Inset: AFM image of the same device. Scale bar is 5 µm. (b) Thermal emission spectra from the quartz-supported device at different bias voltages (from 8 V to 16 V with 2 V step, grey to orange). Smooth lines are the fits obtained using equation 1. The inset shows current–voltage characteristics of the device, as well as the electronic temperatures extracted from the fits on the main panel.

Using equation 1 we obtained reasonable fits to the recorded emission spectra; extracted electronic temperatures and corresponding current–voltage characteristics are presented in the inset of figure 1b. We also evaluated the model for light emission from graphene proposed in [12] but found that the spectra are indistinguishable from grey-body emission in the investigated spectral range. The emission efficiency ($P_{emission}/P_{electric}$) of our devices was ~ $1.6 \cdot 10^{-5}$, obtained using Stefan-Boltzmann law ($P_{emission} = \epsilon A\sigma T^4$, $A$ – device area, $\sigma$ – Stefan-Boltzmann constant) and electrical power ($P_{electric} = IV$) similar to [9, 12]. For the emissivity ($\epsilon$) we used 0.023, following literature [9], the details can be found in the Supplementary information (Section 3).

Our devices demonstrated surprising long-term stability (many hours) at these very high current densities (> $1.7 \cdot 10^8$ A/cm$^2$), comparable with multilayer graphene-based infrared emitters but operated at much lower ($7.1 \cdot 10^6$ A/cm$^2$) current densities [10]. We further discuss a long-term stability and the mechanisms of breakdown in the Supplementary Information (Section 1).

## 3. Enhancement of incandescence using photonic cavity

We prepared another set of hBN/Gr/hBN heterostructures but this time on $SiO_2$/n-Si substrate and carried out the same high current light emission measurements, figure 2a. We found that the emission spectra shift to the visible range and the graphene device incandesces (figure 2b, c). Considering this device is almost identical to that in figure 1, except for the presence of heavily doped silicon substrate, the drastic change in the emission spectra (cf. figure 2c and figure 1b) can be attributed to a resonance effect of a photonic cavity confined by the two interfaces: hBN/air and $SiO_2$/Si, figure 2d.

It is instructive to analyse the parameters of this cavity using a simple model of a Fabry-Pérot resonator [20, 21]. The optical path length of the cavity $D = \sum d_i n_i$ is ~ 740 nm for 290 nm $SiO_2$ and 150 nm hBN/Gr/hBN stack. The refractive indices $n_i$ of $SiO_2$ and hBN show negligible dispersion throughout the visible and near infrared spectral range, averaging at $n_{SiO2}$ ~ 1.49 and $n_{hBN}$ ~ 2.05 [20, 22]. The reflectivities $R$ of the "mirrors" formed by hBN/air and $SiO_2$/Si interfaces, are 0.12 and 0.19, estimated from Fresnel equations $R_{ij} = \left|\frac{n_i-n_j}{n_i+n_j}\right|^2$, where $n_i$ and $n_j$ are the refractive indices of the materials forming interfaces (with $n_{Si}$ ~ 3.8 and $n_{air}$ = 1, from [20]). These relatively low reflectivities lead to the output coupling losses, typically parametrised by the mirror coupling coefficients, $\sum -\log(R) = 3.8$ in our case (2.1 and 1.7, for hBN/air and $SiO_2$/Si mirrors, respectively). The expected photon decay time $\tau = t_{RT}/\sum -\log(R)$ ~ 1.3 fs ($t_{RT} = 2D/c$ the round trip time for the photon in the cavity) matches well the linewidth ($\Delta E_{measured}$ ~ 0.5 eV, cf. $\Delta E = h/2\pi\tau$) of the measured emission peaks of figure 2c (extracted from experimental curves the linewidth values are presented in the figure S7 of the Supplementary Information). Furthermore, the observed peak maxima (~1.6 eV) are very close to the expected second mode of the cavity, $2h/t_{RT} = 1.67$ eV.

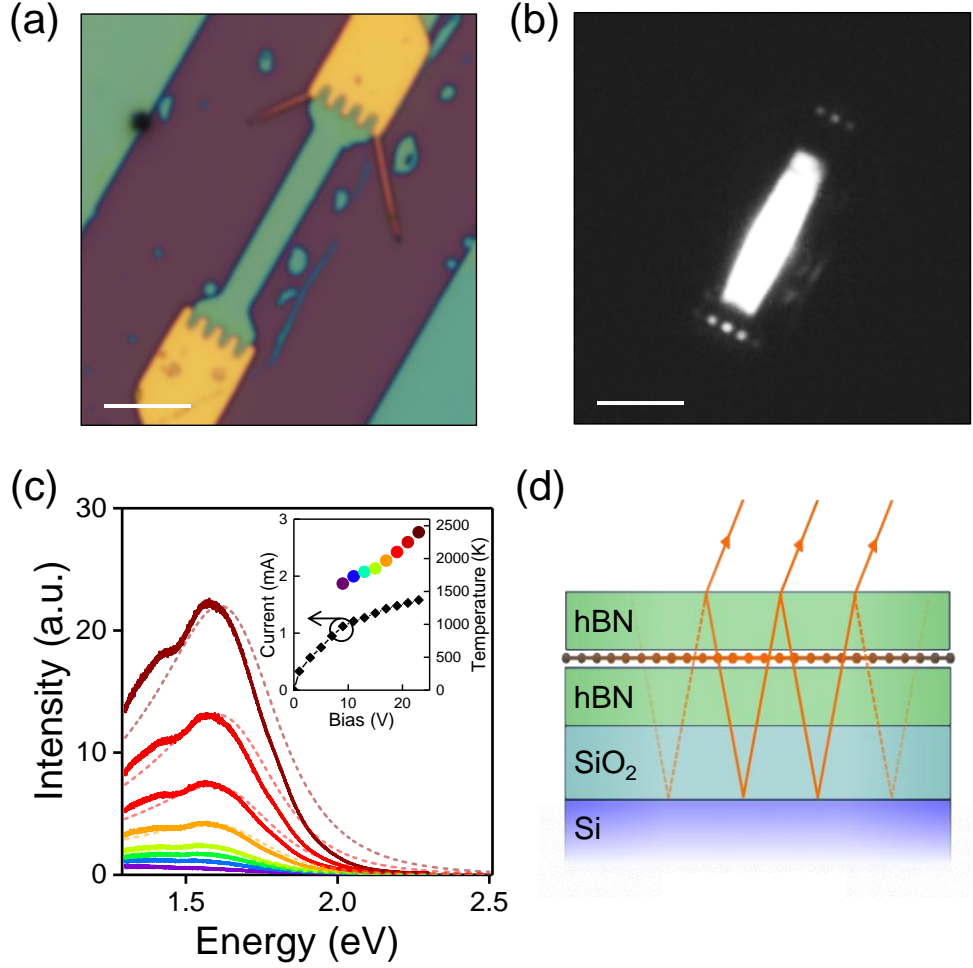

**Figure 2.** Incandescence in ambient conditions of Joule-heated hBN/Gr/hBN/SiO$_2$/Si device. (a) Optical micrograph of hBN/Gr/hBN filament under illumination. (b) Its incandescence image captured by a monochrome camera. Scale bars in (a) and (b) are 5 µm. (c) Incandescence spectra (as well as theory curves, equation 2) of the graphene device at different bias voltages (from 9 V to 23 V in 2 V steps, violet to dark red). The inset shows current–voltage characteristics of the device, as well as the electronic temperatures estimated from photonic cavity model, equation 2. (d) Schematic diagram of a Fabry-Pérot cavity of the air/hBN/Gr/hBN/SiO$_2$/Si system.

We modelled the incandescence spectra by considering the grey-body radiation (equation 1) modulated by a spectral line shape of the second mode of the cavity (using a Lorentzian, $f(h\nu) = \frac{(\hbar/\tau)^2}{4(h\nu - 2h/t_{RT})^2 + (\hbar/\tau)^2}$, following [21]):

$$I_{cavity}(h\nu, T) = I_{grey\ body}(h\nu, T) f(h\nu) \qquad (2)$$

Surprisingly, this simplistic approach provides reasonable quality fits for the data (dashed lines in figure 2c), using temperature as the only fitting parameter. The extracted temperatures are presented in the inset of figure 2c, the values are higher than for the device on quartz (cf. inset of figure 1b) because of the larger electrical power applied in the former case. Our temperatures were somewhat lower (1500–2500 K) in comparison with the emission (2500–2900 K) reported for suspended graphene [12]. We attribute this behaviour to the dissipation of heat to the surrounding media through hBN [11]. The hyperbolic phonon polaritons of hBN [23, 24] are predicted to efficiently (on a picosecond scale) cool hot graphene [11]. To reveal the interplay between the underlying physical mechanisms of cooling one need to perform time-dependent spatially-resolved thermal conductivity measurements, preferably on suspended hBN/Gr/hBN heterostructures, which is beyond the scope of this article.

Another effect of the photonic cavity can be seen in figure 2b – two rows of bright dots on the opposite ends of the device (their positions coincide with the tips of the gold comb electrodes). The gold electrodes are in thermal equilibrium with environment due to their macroscopic size and very good thermal conductivity, hence the temperature would be too low for the incandescence of the gold tips. On the other hand, the incandescent light from the central area of the device could travel through the waveguide formed by a photonic cavity all the way to the contacts, where it radiates possibly via coupling to gold plasmon modes at the contact edges [25]. The waveguide modes guide the graphene light emission over a long distance (at least 5-10 µm) even in this simple cavity, while the performance can be further improved by optimising the dielectric material and thickness to enhance the light emission [26, 27]. The frequency range of the photonic cavity modes can also be adjusted by optimising appropriate geometrical and material parameters.

## 4. Transport measurements in high-bias regime

To better understand the physics of graphene incandescence and to independently estimate the electronic temperatures in a high-bias high-current regime we performed four-terminal transport measurements on an hBN/Gr/hBN Hall bar device, figure 3. Electric field and current density values were obtained from measured 4-probe voltage $V_{xx}$, applied current $I$, and the geometry of the Hall bar ($L \times W$ is $2 \times 1$ µm$^2$). At low biases, the current density increases linearly with the electric field and then deviates from the ohmic regime at $|E| > 0.1$ V/µm, figure 3a. This saturation tendency is typically attributed to the presence of in-plane acoustic and optical phonons [28-30], and other effects, such as Auger scattering [31].

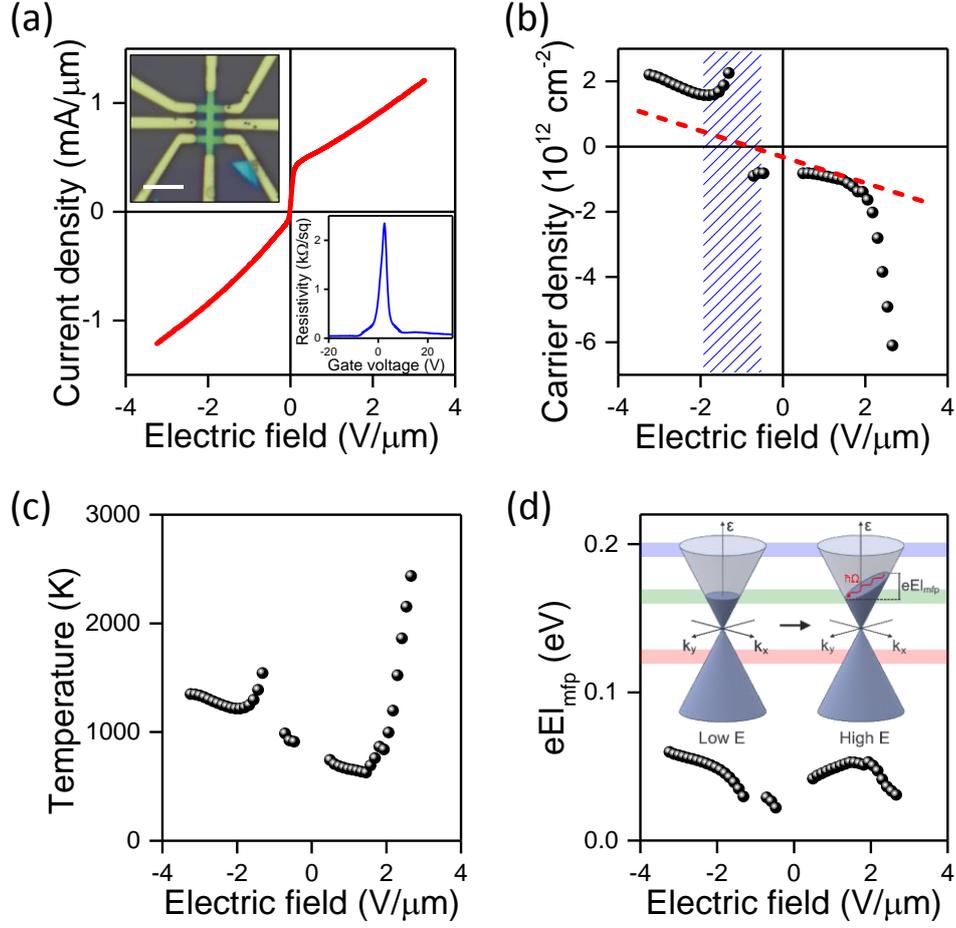

**Figure 3.** Transport characteristics of a multi-terminal hBN/Gr/hBN Hall bar device, measured in a cryostat at 4 K. (a) Current density $j$ vs electric field $E$ at zero gate voltage. Top inset – device micrograph, scale bar 5 µm. Bottom inset – low-bias resistivity vs gate voltage. (b) Charge carrier density vs electric field (at zero gate voltage) measured from Hall effect (black circles), and estimated from self-gating field effect (dashed red line). The divergence at −1 V/µm is due to the change in the carrier type (the blue shaded region). (c) Electronic temperature extracted from data on the panel (b) using equation 3. (d) High electric field and high current induced non-equilibrium Fermi distribution limited by electron-phonon scattering processes. The energy values for in-plane graphene phonons (Γ-point optical phonons – blue line, K-point optical phonons – green line, and K-point transversal acoustic phonons – red line) are from phonopy [32].

To estimate the charge carrier density $n$, we measured our Hall bar device in a perpendicular magnetic field $B$, following the procedure described in [33]. The transversal voltage $V_{xy}$ was linear with $B$ within the measurement range of ±2 Tesla, allowing to estimate $n = -IB/eV_{xy}$ at different electric fields, figure 3b. At a non-zero temperature ($T$) and in the presence of disorder ($\Delta n < 10^{10}$ cm$^{-2}$) the measured carrier density is

$$n = \frac{1}{2}\left(n_g + \sqrt{n_g^2 + 4\left(\sqrt{\left(\frac{\Delta n}{2}\right)^2 + (n_{th}(T))^2}\right)^2}\right) \quad , \tag{3}$$

where $n_{th}(T) = \frac{\pi}{6}\left(\frac{k_B T}{\hbar v_F}\right)^2$ is the thermally-induced carrier density, and $v_F$ is the Fermi velocity of electrons in graphene [34, 35]. The gate induced carrier density $n_g = \frac{\varepsilon\varepsilon_0}{de}\Delta V_g$ ($\varepsilon\varepsilon_0$ and $d$ are the dielectric constant and the thickness of gate dielectric) was estimated from the corrected gate voltage $\Delta V_g$, where the self-gating effect (due to large bias voltage) was taken into account, red dashed line in figure 3b. Using equation 3 we estimated the effective electron temperature of graphene as a function of electric field, figure 3c. The obtained temperatures are in range with those estimated from the emission spectra (figures 1 and 2).

Knowing the current density $j$ and the charge carrier density $n$ we extracted the drift velocity $v_d = j/ne$, and the mobility $\mu = v_d/E$ at different electric fields (and, hence, temperatures). The $v_d$ slowly saturates starting at $E \sim 1.5$ V/µm reaching $4 \times 10^5$ m/s, almost 40 % of the Fermi velocity. The mobility, high initially (~ 100 000 cm$^2$/V·s), drops to below 1000 cm$^2$/V·s at high electric fields. From a simple Drude model we approximated the scattering time $\tau$ and the mean free path $l_{\text{mfp}}$ of the charge carriers from their mobility, $\mu\hbar\sqrt{\pi n}/e = v_F \tau \equiv l_{\text{mfp}}$. For independent scattering processes, the scattering time $\tau$ is dominated by the shortest process, which at high bias fields is usually the emission of the optical phonons [28, 36]. If the potential difference over the characteristic length $l_{\text{mfp}}$ is sufficiently high ($eEl_{\text{mfp}} \geq \hbar\Omega$) a phonon of energy $\hbar\Omega$ can be emitted resulting in the dissipation (heat) of electric energy in graphene lattice, and surrounding hBN and SiO$_2$. This process, in turn, limits the $l_{\text{mfp}}$, and bounds the non-equilibrium Fermi distribution, as shown on the cartoon in figure 3d (depicted as the tilt of the Fermi level). Figure 3d compares the characteristic energy $eEl_{\text{mfp}}$ against the typical (at high density of states in the phonon dispersion spectra) energies of representative phonon branches of graphene [37, 38]. Interestingly, the obtained characteristic energy values ~ 60 meV are sufficiently lower than graphene's in-plane phonon excitation energies, figure 3d. Perhaps low energy acoustic phonons, longitudinal guided modes or other higher order acoustic modes at the interface of hBN and graphene limit the characteristic energy [37]. For a rather thin hBN substrate the characteristic energy might also be bound by optical phonons of SiO$_2$ substrate [28].

## 5. Raman spectroscopy and lattice temperature

To complement the light emission and transport study, we performed Raman spectroscopy on our hBN/Gr/hBN heterostructures. To this end, we employed the same Renishaw spectrometer using 633 nm excitation laser line with incident power of 2 mW. The recorded spectra of the characteristic Raman peaks (G and 2D for graphene, E$_{2g}$ for hBN) for different bias voltages are summarised in figure 4. The intensities of all three peaks monotonically decrease with temperature. There is also a downshift with increasing temperature for all three peaks: 5 cm$^{-1}$, 6 cm$^{-1}$, and 7.5 cm$^{-1}$ for G and 2D peaks of graphene, and E$_{2g}$ peak of hBN at the highest applied bias, respectively. The downshift of the G peak is attributed to the decay of Γ-point optical phonon into acoustic phonons through a range of phonon-phonon scattering processes [39]. Similar anharmonic decay processes (complicated by competing electron-electron scattering) of K-point optical phonons are also responsible for the downshift of the 2D peak of graphene [40, 41]. However, both G and 2D peaks tend to blue-shift as a result of doping [42] and strain [43], thus making it difficult to extract the correct lattice temperature from the Raman peak shift. Furthermore, the linewidth of the 2D peak is also affected by strain, for example, the nanometre-scale strain variations contribute to broadening of the 2D peak [44]. We, therefore, focus below on a linewidth of G peak as the most robust method.

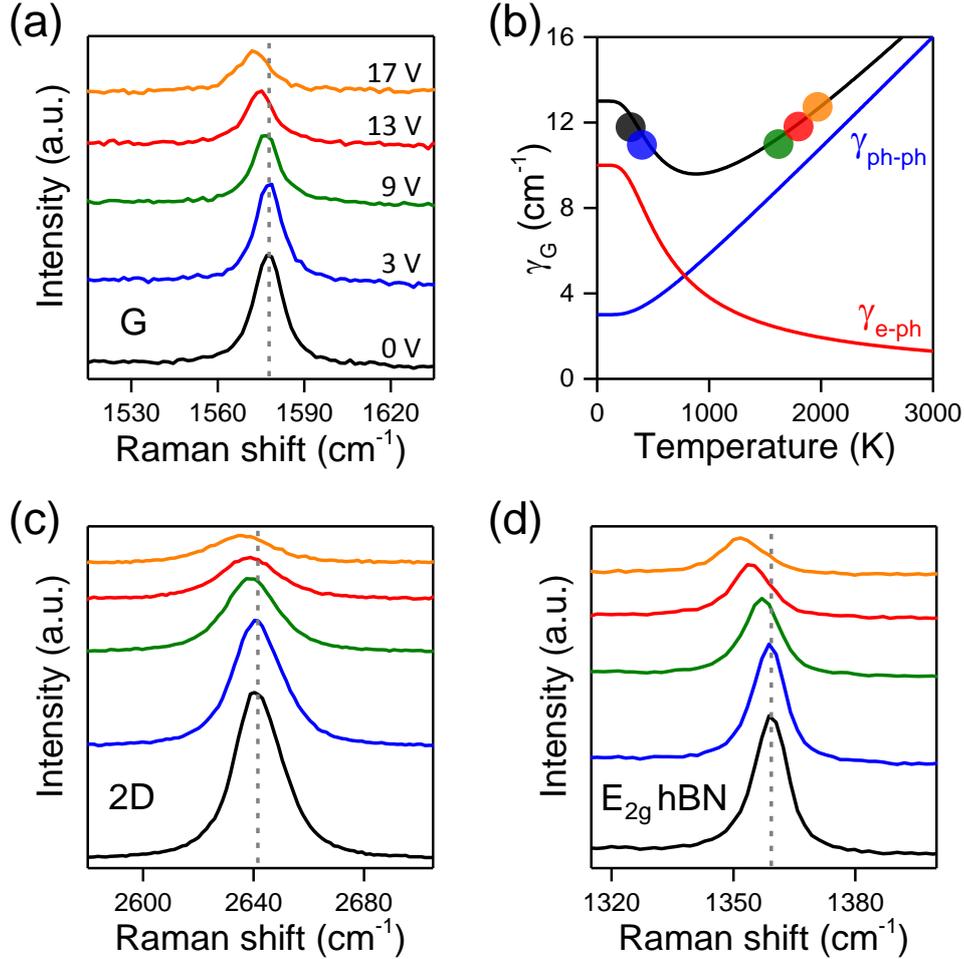

**Figure 4.** Raman spectra of hBN/Gr/hBN device (same as in figure 2) at different bias voltages. (a) Raman spectra of G peak. (b) The linewidth of G peak as a function of temperature from phonon-phonon lifetime broadening (blue line), electron-hole plasma Landau damping (red line), and their sum (black line). The coloured circles (0 to 17 V, black to orange) are the experimental points extracted from the panel (a), plotted vs electronic temperatures extrapolated from emission spectra in figure 2. (c) Raman spectra of 2D peak. (d) Raman spectra of $E_{2g}$ hBN peak.

We observed a non-monotonic behaviour of the linewidth $\gamma(T)$ of the $\Gamma$-point phonon mode (G peak) with temperature, figure 4 a,b. This behaviour is in agreement with previous experimental observations, where linewidth of G peak is also first decreased and then increased with temperature [45]. We follow a simple model, considering two processes by which a $\Gamma$-phonon acquires a finite linewidth $\gamma_G(T)$: by the anharmonic decay into two secondary acoustic phonons, $\gamma_{ph\text{-}ph}(T)$, or by creating an electron-hole pair (Landau damping), $\gamma_{e\text{-}ph}(T,\mu)$ [39, 45]. Here we ignore any higher-order anharmonic coupling processes. The total estimated linewidth of G peak $\gamma_G(T) = \gamma_{e-ph}(T,\mu) + \gamma_{ph-ph}(T)$ is plotted in figure 4b (black line; the anharmonic contribution and the Landau damping are the blue and red lines, respectively). We use 10 cm$^{-1}$ and 3 cm$^{-1}$ for $\gamma_{e\text{-}ph}(0,0)$ and $\gamma_{ph\text{-}ph}(0)$, and include thermally-induced suppression of Landau damping, following procedures from [45, 46]. The $\gamma_G$ values from figure 4a plotted as coloured circles vs electronic temperatures extracted from figure 2c, are in a good agreement with the simple model suggesting the dominant anharmonic scattering mechanism.

Lastly, the hBN E$_{2g}$ peak is also downshifting and broadening with temperature (figure 4d) due to similar anharmonic phonon decay processes in hBN [47]. This suggests there is a strong heat transfer from graphene to the encapsulating hBN. Despite the heat dissipation in the encapsulating hBN layers, the electronic temperatures of graphene were well above 2000 K, as estimated from the incandescence spectra. Understanding the detailed mechanism of heat dissipation through the encapsulating hBN requires further theoretical investigations.

## 6. Conclusions

We demonstrated an interesting region of a parameter space (high current and electric field, high electronic and lattice temperatures), accessible in simple graphene/hBN heterostructures (and in the majority of existing devices) in ambient conditions. Electrically tunable light emission of graphene is modified by a photonic cavity formed at the interfaces, which can be further improved by optimising the parameters of the cavity. This atomically-thin light source is integrated with the cavity, which forms a waveguide, paving the way for new types of electrically-coupled photonic integrated circuits. This level of integration can be brought even further, by combining the photonic circuit elements with electrically-driven phononic and mechanical modes. We believe our study will stimulate further experimental and theoretical work in this fascinating research direction, where many interesting questions are now open, such as the understanding the mechanisms of heat dissipation in these devices, including the effects of hyperbolic phonon polaritons in hBN.

**Acknowledgments**

This work was supported by the European Research Council, the EU Graphene Flagship Program, the Royal Society, the Air Force Office of Scientific Research, the Office of Naval Research and ERC Synergy Grant Hetero2D. Servet Ozdemir acknowledges PhD studentship from EPSRC funded Graphene NOWNANO CDT. Artem Mishchenko acknowledges the support of EPSRC Early Career Fellowship EP/N007131/1.

# Graphene hot-electron light bulb: incandescence from hBN-encapsulated graphene in air

Seok-Kyun Son[1,4], Makars Šiškins[2,4], Ciaran Mullan[2,4], Jun Yin[2], Vasyl G. Kravets[2], Aleksey Kozikov[1], Servet Ozdemir[2], Manal Alhazmi[2], Matthew Holwill[1], Kenji Watanabe[3], Takashi Taniguchi[3], Davit Ghazaryan[1], Kostya S. Novoselov[1,2], Vladimir I. Fal'ko[1,2], Artem Mishchenko[1,2,5]

[1]*National Graphene Institute, University of Manchester, Manchester M13 9PL, UK*

[2]*School of Physics and Astronomy, University of Manchester, Manchester M13 9PL, UK*

[3]*National Institute for Materials Science, 1-1 Namiki, Tsukuba, 305-0044, Japan*

[4] These authors contributed equally

[5] e-mail: artem.mishchenko@gmail.com

1. **Device performance and stability in high current regime**

We have measured more than a dozen of different devices, both standard multi-terminal transport devices and specifically designed filaments, and most of them survived long-term measurements of at least several hours at current densities in excess of $10^8$ A/cm$^2$. Naturally, with the higher current densities the life-time of the devices drops exponentially. As an example, figure S1 shows the stability of a conventional Hall bar device (hBN/graphene/hBN) in a high current regime – we managed to record three breakdown curves at ~ 1, 2 and 3·$10^8$ A/cm$^2$ current densities, with the life-time of the order of $10^5$, $10^3$, and $10^1$ seconds, respectively.

For the standard Hall bar devices, the main mechanism of failure is most likely the breakdown at the graphene/gold interface probably due to an increased mobility of gold atoms at elevated temperatures in the presence of high electric fields, which exposes the interface to the ambient and allows for thermal oxidation of graphene along the 1D graphene/metal contact. In the specially designed filaments (with a wide contact area and a long comb shaped contact perimeter) the breakdown usually happens after much longer periods of time, because the graphene/metal interface is at much lower temperatures thanks to the special design of the contact (cf figures S1d and S2a). Typically, the device breaks in the middle of the filament (figure S2) and also at higher current densities (> 3·$10^8$ A/cm$^2$). The location of the break usually can be traced back to the presence of bubbles of trapped contaminants between the graphene layer and one of the encapsulating hBN crystals. At these high temperatures the content of the bubbles (mostly hydrocarbons) decomposes producing reactive chemical species which can corrode the graphene (as seen in figure S2 panel d), or even blast the heterostructure due to the formation of gaseous products from thermal decomposition of trapped contaminants (figure S2 panel b).

The maximum attainable current density is comparable to non-encapsulated devices measured in vacuum, $4.5\cdot 10^8$ A/cm$^2$ [1], and approximately twice smaller than that of single-walled carbon nanotubes on SiO$_2$ substrate, $10^9$ A/cm$^2$ [2]. The hBN protection properties were recently exploited for infrared (~ 4 μm) emitters based on a few (6-8) layer graphene [3]. Encapsulation of the few-layer graphene heating element with 13 nm hBN allowed extended operation (> 1000 hours) in ambient conditions. The device was Joule-heated (current density < $10^7$ A/cm$^2$) to 440-530 K using AC current. Thus, the long-term stability of hBN encapsulated devices depends on several factors, such as the presence of trapped air-born hydrocarbons in the form of bubbles during the device fabrication, the current density, the contact resistance, etc.

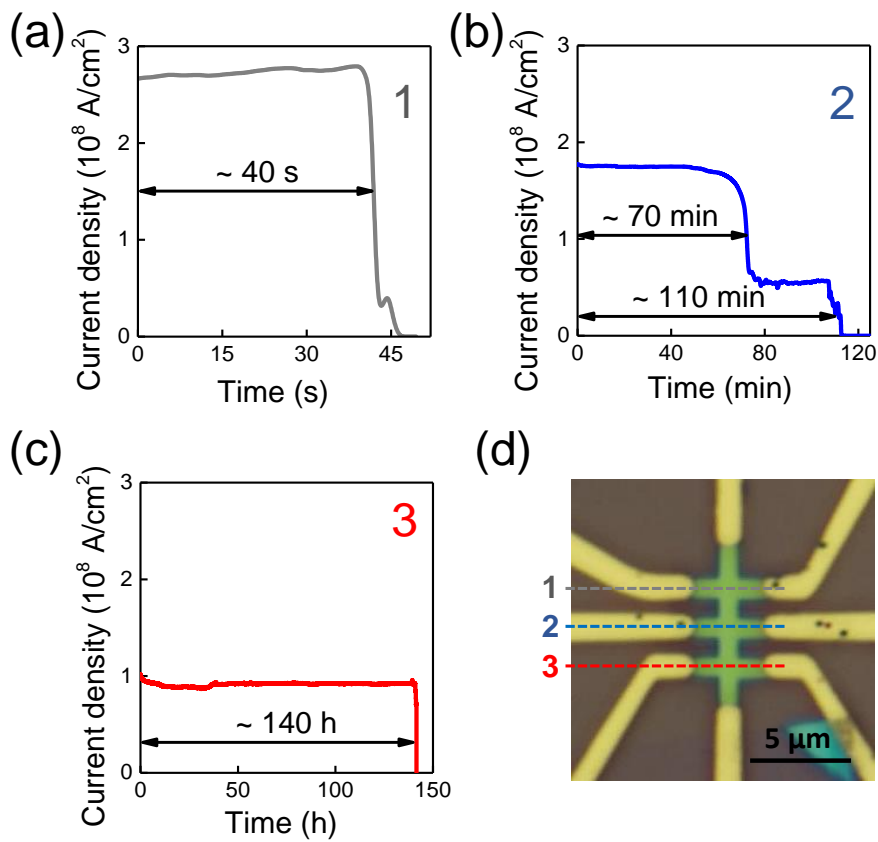

Figure S1. Long-term stability of standard Hall bar devices (encapsulated hBN/graphene/hBN heterostructures) in a high current regime in ambient conditions.

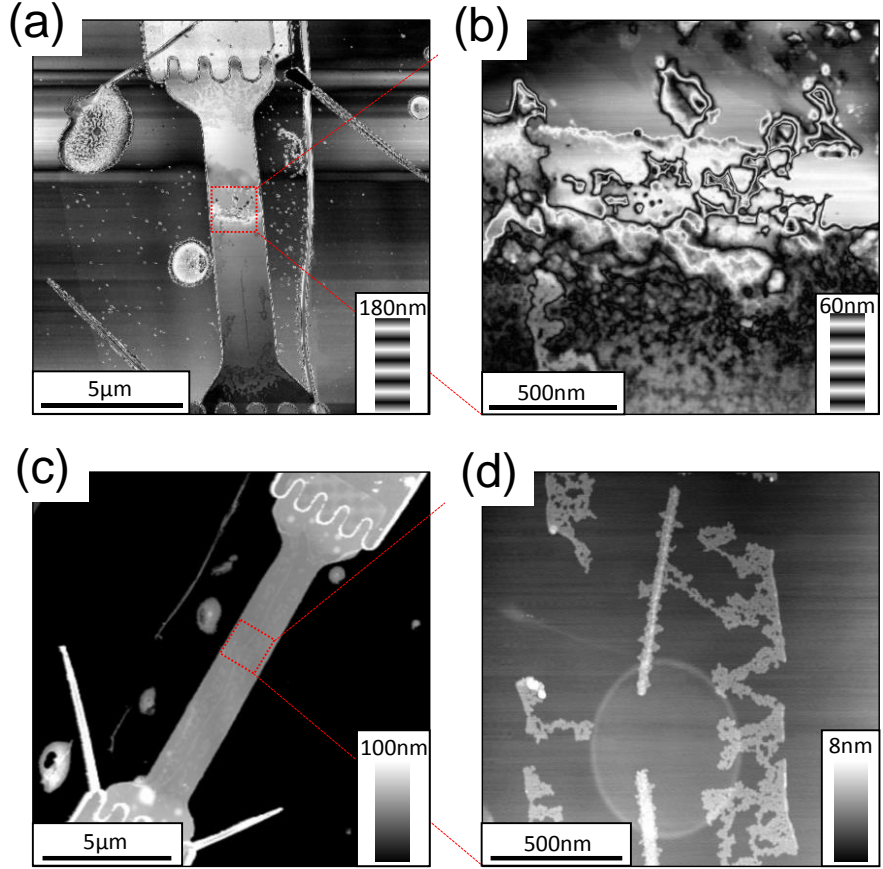

**Figure S2.** AFM images of hBN-encapsulated graphene filaments after total (panels a and b) and partial (panels c and d) breakdown. Device on panels a and b has burned after glowing for ~ 12 hours at ~ $2\cdot10^8$ A/cm$^2$ current density. Device on panels c and d (the same device as in the figure 2, main text) developed a blister when operated at ~ $3\cdot10^8$ A/cm$^2$ for a couple of hours. Raman measurements confirm that graphene is partially decomposed: intensity of 2D peak in the blister area is 10 times smaller as compared to the rest of the device. Nevertheless, even after the formation of blister and partial breakdown of the graphene the device was still operational and survived many high current regime measurements afterwards. The corrugated features present in device channel area (most notable on the panel d) are the polymer residues on top of hBN left after the device fabrication.

## 2. Extracting temperatures from grey-body radiation and Fabry-Perot cavity models

Spectra of thermal radiation for an hBN-encapsulated graphene can be approximated by Planck's law, modified by emissivity $\epsilon$ (grey-body radiation):

$$I_{grey-body}(h\nu, T) = \epsilon \frac{2(h\nu)^3}{h^2 c^2} \frac{1}{\exp\left(\frac{h\nu}{k_B T}\right) - 1}$$

Following Kirchhoff's law of thermal radiation we assume the emissivity of monolayer graphene from its absorptivity $\epsilon = \alpha \approx 0.023$, which, in the first approximation, is energy-independent in a spectral range from visible to near infrared [4]. To fit the experimental data, we also used an additional proportionality coefficient, $k$, to grey-body model to normalize on arbitrary units of spectra acquired by Si-based CCD detector: $I_{measured}(h\nu, T) = k \times I_{grey-body}(h\nu, T)$. In order to simplify the fitting equation, we hereby absorbed all multiplicative constants as a single proportionality coefficient, $K = \frac{2k\epsilon}{h^2 c^2}$, so that $I_{measured}(h\nu, T) = K \times (h\nu)^3 \left(\exp\left(\frac{h\nu}{k_B T}\right) - 1\right)^{-1}$. Subsequently, we deducted $K$ from the fit of the highest-intensity spectrum (acquired at 16 V) and then fixed this value, simplifying the fitting procedure to a single parameter (temperature, $T$) fit for all subsequent iterations. As seen in figure S3, panels a and b, for hBN/graphene/hBN heterostructures on quartz substrate light emission spectra closely follow the grey-body model.

In contrast, when SiO$_2$/Si substrate is used, a photonic cavity formed between hBN/air and SiO$_2$/Si interfaces shifts the emission spectra to the visible range. Consequently, for the following case we considered the incandescence model modulated by a spectral line shape of the second mode of the cavity, using a Lorentzian:

$$I_{measured}(h\nu, T) = K \times \frac{(h\nu)^3}{\exp\left(\frac{h\nu}{k_B T}\right) - 1} \times \frac{(\hbar/\tau)^2}{4(h\nu - 2h/t_{RT})^2 + (\hbar/\tau)^2}$$

Since both photon decay time, $\tau$ and round trip time, $t_{RT}$ could be estimated from known dimensions of the cavity, model significantly simplifies to have a temperature as the only fitting parameter (see figure S3, panels c and d).

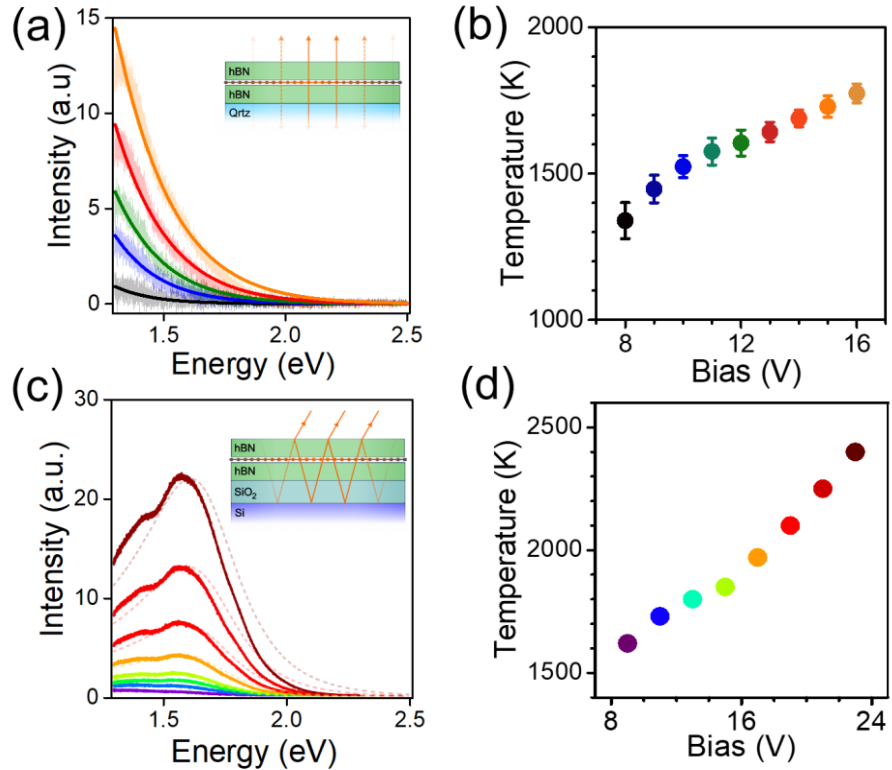

**Figure S3.** Thermal radiation of Joule-heated hBN/graphene/hBN devices. (a) Thermal emission spectra from the quartz-supported device at different bias voltages (from 8 V to 16 V, grey to orange). The inset shows schematics of light propagation inside the heterostructure. (b) Temperatures deducted

from the fit. (c) Spectra from the SiO$_2$/Si-supported device at different bias (from 9 V to 23 V, violet to dark red). The inset: schematics of Fabry-Perot cavity formed. (d) Temperature values extracted from the fit.

Remarkably, this approach provides a fair estimate of sample temperatures under a range of applied electric fields. Several characteristics of the spectra, however, such as unexpected shoulder seen at ~ 1.3 eV in figure S3c, remained yet unclear and required a more sophisticated analysis for a better understanding of the effect of photonic cavity. We described a more involved model of this particular case in details in Section 4.

### 3. Incandescence efficiency – thermal radiation vs electrical power

A supplied electrical power $P_{electrical} = I \cdot V$ dissipates as heat, which raises the temperature of a device (figure S4a). The total radiated power can be approximated from the temperature of incandescence using Stefan-Boltzmann law, $P_{optical} = A\epsilon\sigma T^4$ (figure S4b). The efficiency of light emission ($P_{optical}$ / $P_{electrical}$) of our devices was in the range of 1–3·10$^{-5}$, therefore only a small fraction of supplied electrical power is radiated away. This suggests that hBN encapsulation, substrate, and the metal contacts sink most of the released heat. When directly comparing with commercial incandescent lamps, the efficiency of our graphene-based filaments is ~ 2·10$^{-5}$, while commercial light-bulbs reach values about 2·10$^{-2}$ [5]. This would correspond to three orders of magnitude better radiation efficiency for conventional macroscopic light-sources. However, we believe such comparison of macroscopic alternatives with the 2D microscopic light emitters may appear ambiguous and shall be better compared with other known sources of a more commensurate scale. In figure S4c we compare the efficiency of thermal radiation of our devices with those reported in literature. The efficiency of our devices was an order of magnitude better than that for graphene on SiO$_2$/Si substrate [6, 7], and comparable to hBN encapsulated multilayer graphene thermal emitters [3]. By suspending the graphene filament, the efficiency can be increased by 2–3 orders of magnitude [8], which is due to diminished heat transfer to the substrate. The reason for better efficiency of hBN encapsulated devices (compared to graphene-on-SiO$_2$ devices) remains unclear and requires further investigations.

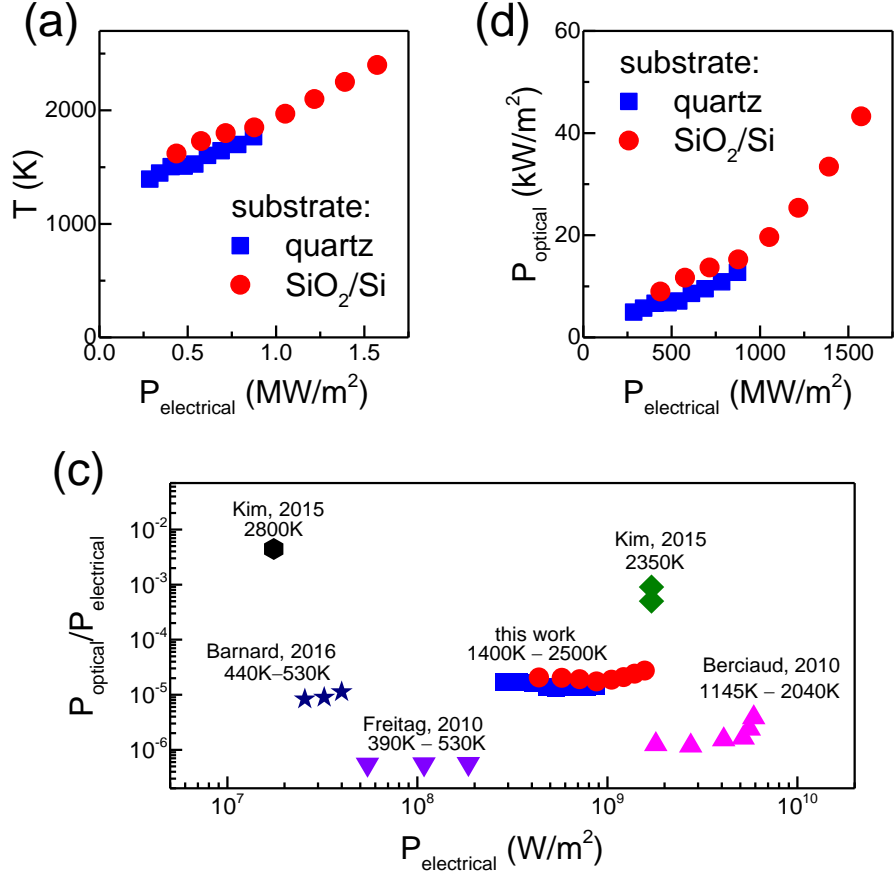

Figure S4. Comparison of attainable temperatures (panel a), optical power (panel b) and emission efficiency (panel c) versus supplied electrical power.

## 4. Thermal radiation in a photonic cavity

Here we consider the following 5-layer structure (figure S5) for a black-body-like thermal radiation emitter. The emitter is represented as the thin layer of hBN dielectric of thickness $d_1 = 30$ nm (medium 1), followed by a graphene monolayer, $d_2 = 0.34$ nm (medium 2), then another thin layer of hBN dielectric of thickness $d_3 = 50$ nm (medium 3), then a $SiO_2$ dielectric of thickness $d_4 = 290$ nm (medium 4) on a semi-infinite doped silicon substrate (Si).

In our modelling, we estimate the emission from graphene monolayer as the standard black-body radiation described by Planck's formula. Following Landau and Lifshitz approach [9] the radiation energy intensity $U(\omega,T)$ (power per unit area at unit frequency interval) can be expressed as [10]:

$$U(\omega, T) = \frac{\hbar\omega}{e^{\hbar\omega/k_B T}-1} \frac{n_{gr}^2 \omega^2}{(2\pi)^3 c^2}, \quad (1)$$

where $T$ is a black-body thermal equilibrium temperature, $\hbar$ and $k_B$ are the reduced Planck's constant and the Boltzmann's constant, respectively, $n_{gr}$ is the real part of the graphene refractive index, $c$ is the speed of light, and $\omega$ is spectral radiation frequency.

We consider graphene as the source of black-body radiation at a given temperature, *T*. The emitted radiation is reflected by the photonic cavity, and therefore the radiation leaving the very top surface of the heterostructure consists of directly emitted and reflected components. In the case of an isotropic black-body-like emitter, we have a contribution in measured intensity from light emission directly from the top surface forwarded to the photodetector and the part modified by the photonic cavity.

The intensity of the emission from the graphene surface (figure S5) is determined by the integration of all the radiation rays originated from it, by analogy with a radiative heat transfer [11, 12]:

$$I(\omega) = \int U(\omega, T) S(\omega, \beta) \beta d\beta, \quad (2)$$

where the Planck's black-body distribution function $U(\omega,T)$ is weighted by $S(\omega,\beta)$ over wavevectors $\beta$. The weighting function includes reflections from interfaces for both *p*- and *s*-polarised light:

$$S(\omega, \beta) = 0.5 \left( R_p(\omega, \beta) + R_s(\omega, \beta) \right), \quad (3)$$

where $R_{p,s}(\omega, \beta) = |r_{p,s}(\omega, \beta)|^2$ are the *p*- and *s*-polarized light reflectivities from the photonic cavity in a direction perpendicular to interfaces in the heterostructure [13, 14]. In turn, the dispersion relation for multiple *p*- and *s*- polarized waves was derived using the Fresnel amplitude reflection coefficients [13]:

$$r_{p,s}(\omega, \beta) = \frac{r_{j,j+1}^{p,s} + r_{j+1,j+2}^{p,s} \exp(2iq_{j+1}d_{j+1})}{1 + r_{j,j+1}^{p,s} r_{j+1,j+2}^{p,s} \exp(2iq_{j+1}d_{j+1})} \quad (4)$$

Finally, the Fresnel reflection coefficients for *p*- and *s*- polarization at the interfaces are given as

$$r_{j,j+1}^s = \frac{q_j - q_{j+1}}{q_j + q_{j+1}} \quad (5a)$$

$$r_{j,j+1}^p = \frac{\varepsilon_{j+1}(\omega) q_j - \varepsilon_j(\omega) q_{j+1}}{\varepsilon_{j+1}(\omega) q_j + \varepsilon_j(\omega) q_{j+1}} \quad (5b)$$

Here $q_j = \sqrt{\varepsilon_j(\omega) \omega^2/c^2 - \beta^2}$ and $\beta = (\omega/c) \sin\theta$ are the components of the wavevectors $k_j$ perpendicular and parallel to the surface of the layers, *j* is a layer index.

The dielectric functions $\varepsilon_1(\omega) = \varepsilon_3(\omega)$, $\varepsilon_2(\omega)$, $\varepsilon_4(\omega)$, and $\varepsilon_5(\omega)$ for top and bottom hBN layers, graphene, SiO$_2$ and Si, respectively, were used to calculate the Fresnel reflection coefficients $R_{p,s}(\omega,\beta)$. The optical constants of SiO$_2$ film and Si substrate were parameterized by the Cauchy and Cauchy-Urbach functions, respectively [14]. To find Cauchy coefficients we carried out spectroscopic ellipsometry measurements of SiO$_2$ film ($d_2$ = 290 nm) on top of Si substrate. The hBN refractive index was also modelled by Cauchy dispersion law using ellipsometric data taken from [6]. We considered only three coefficients $A = 1.97$, $B = 0.6\times10^{-3}$ μm$^{-2}$, $C = 1.9\times10^{-3}$ μm$^{-4}$ for the perpendicular component of the refractive index [15]. The corresponding refractive indices of SiO$_2$ and hBN layers show low dispersion throughout the visible and near infrared spectral range, averaging at $n_{SiO2} = 1.49\pm0.015$ and $n_{hBN} = 2.05\pm0.1$.

The dielectric function of a monolayer graphene can be obtained from the optical conductivity [16]:

$$\varepsilon(\omega) = 1 + \frac{i\sigma(\omega)}{1\pi\omega\varepsilon_0 t_{gr}}, \tag{6}$$

where $t_{gr} = 0.34$ nm is the thickness of graphene, and $\varepsilon_0$ is the vacuum permittivity. The optical conductivity $\sigma(\omega)$ is dominated by the two terms, the intraband electron-photon scattering processes, $\sigma^{intra}(\omega)$, and the direct interband electron transitions, $\sigma^{inter}(\omega)$ [16]. These two terms are as follows:

$$\sigma^{intra}(\omega) = \frac{ie^2 k_B T}{\pi^2 \hbar^2 (\omega + i/\tau)} \ln\left[2\cosh\left(\frac{E_F}{2k_B T}\right)\right] \tag{7a}$$

$$\sigma^{inter}(\omega) = \frac{e^2}{4\hbar}\left\{\frac{1}{2} + \frac{1}{\pi}\tan^{-1}\left(\frac{\hbar\omega - 2E_F}{2k_B T}\right) - \frac{i}{2\pi}\ln\left[\frac{(\hbar\omega - 2E_F)^2}{(\hbar\omega - 2E_F)^2 + 4(k_B T)^2}\right]\right\} \tag{7b}$$

Here, $E_F$ is the Fermi level of graphene, which is dependent on the charge carrier density, and $\tau$ is the carrier momentum relaxation time.

Using equations 4-7 we calculated the polarized reflectivity at the following interfaces: graphene/hBN, hBN/SiO$_2$ and SiO$_2$/Si, and then computed the total reflection coefficients $R_{p,s}(\omega,\beta)$. The spectral intensity of emission, $I(\omega)$, equation 2, was evaluated by the integration over $\beta$ (over all directions at which emission energy leave a unit area of graphene surface) from 0 to $k_0 = 2\pi/\lambda$ ($k_0$ is the wavevector of photon in the air) or, or alternatively, by transforming the integral to a variable polar angle $\theta$ and integrating over $\theta$ between 0 and $\pi/2$.

Figure S6 shows the computed intensities of the emission from the modelled system for different temperatures, as a function of the photon energy. For comparison, the experimental spectral emission is also displayed in figure S6. The modelled (at $T = 2400$ K) and the experimental (black line) spectra match quite well (figure S6b). Note that the small feature (shoulder peak) at around 1.4 eV on experimental curve is also partially reproduced on the theoretical curve (but not on the simple Fabry-Perot model as in section 2 of the Supplementary, or in the main text). Therefore this feature can be attributed to the presence of additional interfaces (such as hBN/SiO$_2$) included in the model in this section, although this would require additional confirmation by measuring a range of devices with different thickness of hBN and SiO$_2$ dielectric layers.

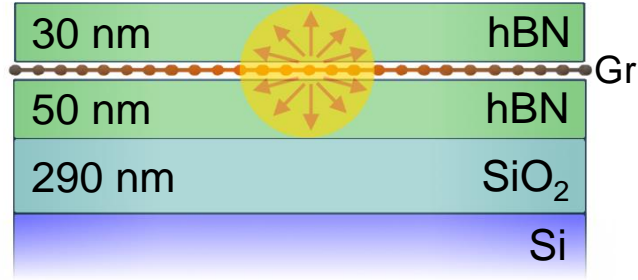

Figure S5. Schematics of the considered structure consisting of single layer graphene, encapsulating hBN stacks, SiO$_2$ layer and Si$^+$ substrate.

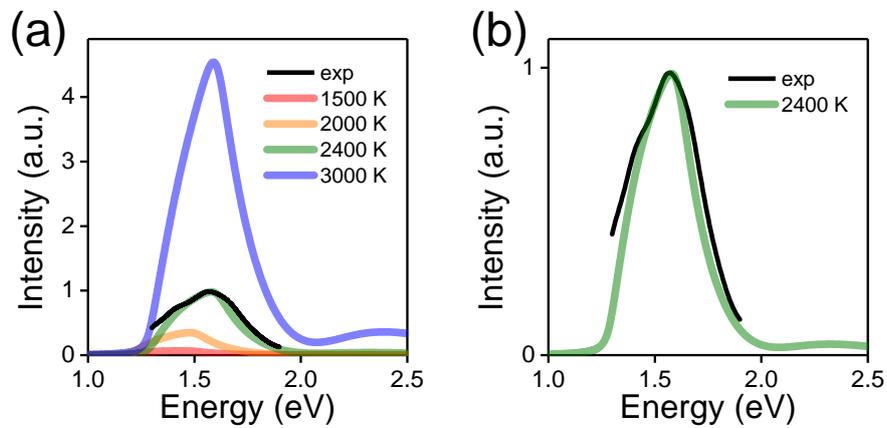

Figure S6. Computed emission from Si/SiO$_2$ supported hBN/graphene/hBN heterostructure as a function of photon energy for different temperatures. For the modelling of thermal radiation we used the structure shown in figure S5.

## 5. Linewidth of the incandescent spectra from the photonic cavity

It is interesting to compare the linewidth of the light emission spectra from the photonic cavity with the photon decay time estimated from the known parameters of the cavity. Figure S7 shows the linewidth extracted from the Lorentzian peak fits for spectra obtained at different bias voltages (and, hence different electronic temperatures). The linewidth is mostly independent of applied bias, except for the lowest bias case, where it slightly deviates from the average value. The average linewidth is ≈ 0.5 eV, in a good agreement with the expected photon lifetime in our photonic cavity.

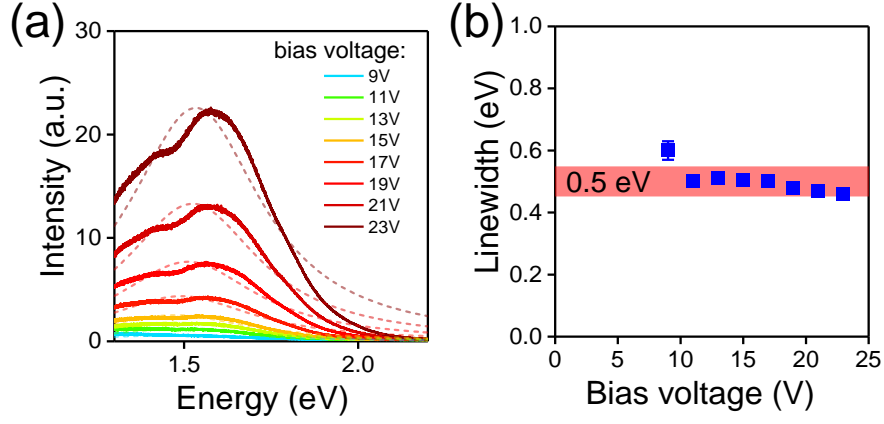

Figure S7. Linewidths of the light emission spectra (extracted from Lorentz peak fits, dotted curves on panel a) vs bias voltage (panel b).

## 6. Temperature distribution in hBN/graphene filaments

To characterise the thermal management in our hBN/graphene hot filaments we considered a simple model [3, 8, 17] based on a 1D heat diffusion equation

$$\nabla(k\nabla T) + \frac{p}{lwd} - \frac{2g}{d}(T - T_0) = 0 \qquad (8)$$

Here, $T$ is a function of $x$ coordinate along the graphene channel of length $l$, width $w$, and thickness $d$. We choose $x = 0$ to be in the middle of the filament and the metal contacts are located at $x = \pm l/2$. $\frac{p}{lw} \equiv P$ is the electric power (Joule heating) per unit area (W·m$^{-2}$), $k$ is the in-plane thermal conductivity (W·m$^{-1}$·K$^{-1}$) and $g$ is an out-of-plane (interfacial) thermal conductance (W·m$^{-2}$·K$^{-1}$). $T_0 = 300$ K is the ambient temperature of a substrate, contacts and air. In a zeroth approximation, we assume that $P$ is uniform along the channel, and both $k$ and $g$ are independent of temperature. Then, with standard boundary conditions, $T(\pm l/2) = T_0$, the solution of equation 8 is straightforward:

$$T(x) = T_0 + \frac{P}{2g}\left(1 - \frac{\cosh(mx)}{\cosh\left(m\frac{l}{2}\right)}\right), \qquad (9)$$

where $m = \sqrt{\frac{2g}{kd}}$. The temperature profiles from equation 9 are summarised in the figure S8, for a range of $k$, $g$, and $P$. The best match to our data (power vs temperature) was for $g \sim 3.5 \cdot 10^5$ W·m$^{-2}$·K$^{-1}$ and $k \sim 500$–$2500$ W·m$^{-1}$·K$^{-1}$. The overall shape of the temperature profile is rather insensitive to the value of $k$, as long as $g$ is above $10^5$. The smaller the $k$ the more flat is the temperature profile. In our simulations we ignored the temperature dependence of $k$ (usually, $k(T)$ is modelled as $k(T) = k_0 \left(\frac{T_0}{T}\right)^\gamma$, where $k(T_0) = k_0$ and $\gamma \sim 1.9$ [17]), the main effect of $k(T)$ of this form is to produce a "hot spot" in the middle of the device, the effect which we did not observe in our case.

The estimated thermal conductance $g$ is well above the measured conductance between graphene and air (2.9·10⁴ W·m⁻²·K⁻¹), and even exceeds the theoretical limit from the kinetic theory of gases [18]: $g_{max} \approx \frac{5}{8} P_{air} \sqrt{\frac{3k_B T}{m}} \sim 10^5$ ($P_{air}$ is the atmospheric pressure, and $m$ is the molecular weight of air). On the other hand, the obtained value of $g$ is ~ 20 times smaller than that reported for graphene/hBN, and approximately two orders of magnitude lower than reported for graphene/SiO$_2$ and graphene/metal interfaces [19, 20]. We believe the hBN encapsulation provides an efficient heat spreader, which prevents the formation of the "hot spot" and allows for a higher current densities and electrical power to be applied to our devices. We would like to note that the effect of heat spreading was recently reported for hBN films and laminates [21, 22].

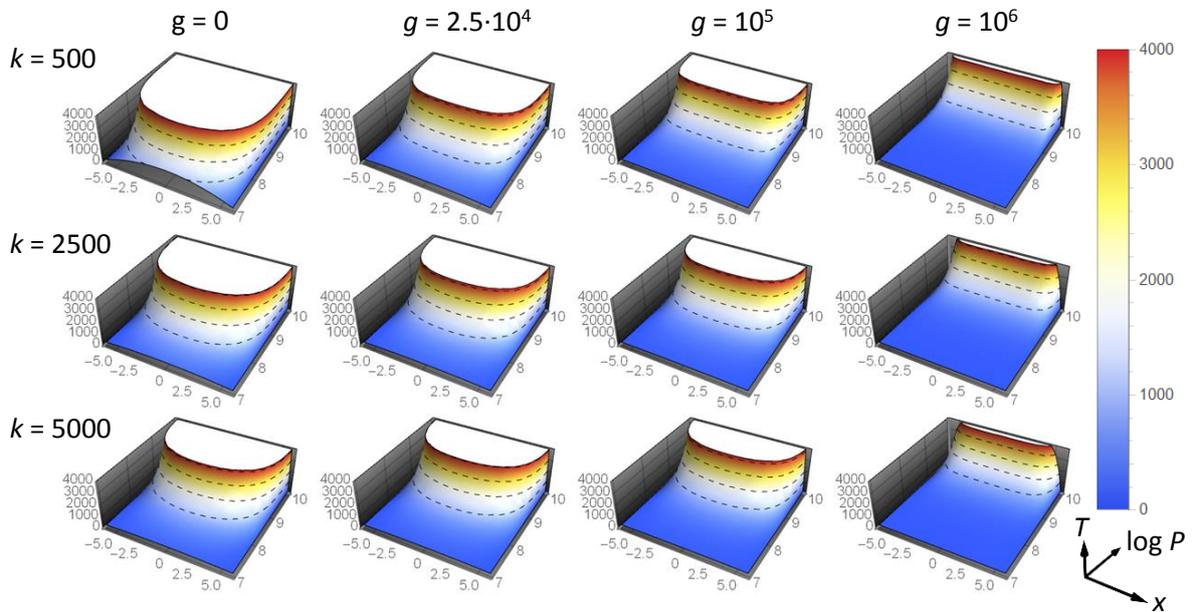

Figure S8. Temperature profiles along the graphene filament (*x*-axis) calculated for different in-plane (*k*) and out-of-plane (*g*) heat transfer contributions.

## 7. Possible plasmonic emission from the metal contacts

Here we would like to discuss the possible explanations for the presence of the light emission spots at the tips of the comb shaped gold contacts (as on figure 2b in the main text). The direct thermal emission due to the local heating at the graphene/gold interface is rather unlikely (although cannot be completely excluded) due to a very good thermal conductivity of graphene/gold interface [20] and a low contact resistance of 1D contacts between hBN/graphene/hBN stack and metal electrode [23]. On the other hand, we are able to account for the bright dots in the vicinity of gold electrodes using a simple cavity model (inset in the figure S9) which predicts a plasmon enhancement of the thermal emission. We consider a light propagation in the dielectric cavity under average angle ($\theta = 45°$), because the large fraction of the emitted light at this angle can be trapped into a photonic cavity and can propagate a relatively long distance. The transmission coefficients for the cavity can be obtained by the following reasoning: the transmitted field is the sum of the first reflection at the SiO$_2$/Si interface and the second reflection at the hBN/air interface, and the cavity will collect the light around the metal contacts after $N = L/(4d)$ consecutive reflections ($L \approx$ 2-5 µm is the distance between the glowing hot part of the device and the metal contact, and $d$ is the thickness of the cavity).

The corresponding p-polarised spectra (which can excite a plasmon on a surface of gold electrodes) propagating along the dielectric cavity are shown in figure S9. Black curve displays the p-polarised emission spectra, $E_p$, in the cavity after two consecutive interferences in vicinity of graphene. Red curve is the intensity of p-polarised thermal radiation in the vicinity of electrodes, which can be estimated as $E_p^N$. Blue curve is the plasmon-enhanced spectrum due to the coupling of the incoming light to the plasmon mode of gold. In this case the plasmon mode will resonate in the exciting field and enhance the local electromagnetic field by a factor $Q \sim \mathrm{Re}(\varepsilon(\omega))/\mathrm{Im}(\varepsilon(\omega))$, (where $\varepsilon(\omega)$ is the complex dielectric function of a 50 nm gold film extracted from ellipsometric measurements). In addition to the surface plasmons of the thin gold field the surface roughness of gold can also couple light to plasmons at "hotspots" [24]. Interestingly, the plasmon resonance slightly shifts the thermal emission spectra in vicinity of Au electrodes to a longer wavelength (cf. blue and red curves in figure S9).

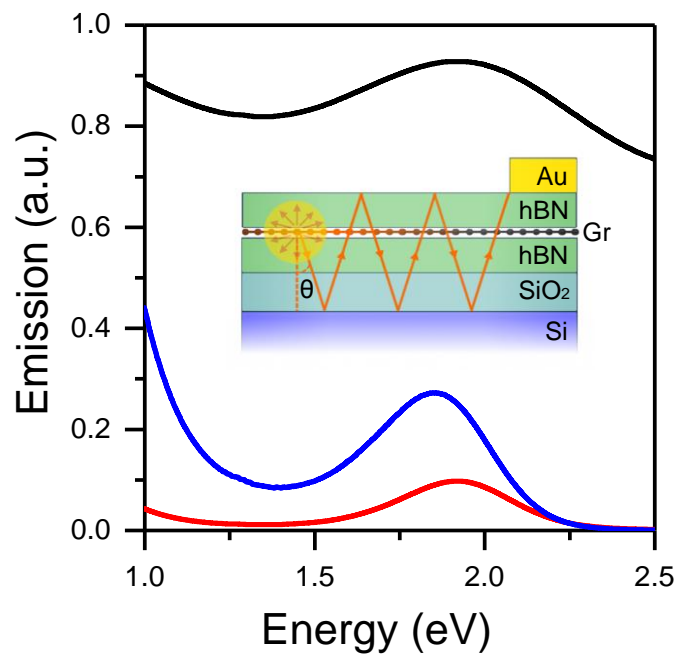

Figure S9. Effective spectral emission from hBN/Gr/hBN heterostructure propagating along the dielectric cavity towards the gold electrodes: black curve – intensity of p-polarised light around graphene-based nanostructure; red curve – reduced thermal emission in the vicinity of Au electrodes for the same polarisation; blue curve – plasmon enhancement of the electromagnetic field at the surface of Au electrodes.

## 8. Carriers density: self-gating effect, thermally induced carrier density and disorder

The charge carriers can be distributed non-uniformly due to a disorder, such as electron-hole puddles, the presence of a temperature gradient and a self-gating at large bias voltages. Compared to $SiO_2$, the hBN substrate reduces the disorder in graphene to a residual value of ~ $10^{10}$ cm$^{-2}$ [25, 26]. We confirm the low density of defects and disorder by transport measurements and Raman spectroscopy. For example, the linewidth of Raman 2D peak was < 20 cm$^{-1}$ implying mobility > 10000 cm$^2$V$^{-1}$s$^{-1}$ at room temperature [27]. The low-temperature mobility was ~ 100000 cm$^2$V$^{-1}$s$^{-1}$ (from transport measurements). The high mobility can be traced back to a low level of disorder [27, 28]. The most likely cause of a disorder is strain inhomogeneity on a sub-micron length scale [29]. At the same time, this low level of disorder does not exclude the possibility of a finite doping due to the presence of impurities during fabrication procedure as well as exposure to light and high temperatures during device operation. For example, a laser-induced doping >$10^{12}$ cm$^{-2}$ was achieved in hBN/graphene heterostructures at a low level of disorder (<$10^{11}$ cm$^{-2}$) by Neumann et al. in [30]. Nevertheless, although significant at low temperatures, the effect of disorder completely vanishes above 1000 K (figure S10).

Besides disorder, the self-gating effect leads to non-uniformity of charge distribution, especially at large bias voltages, $V_b$. Self-gating effect can be understood as follows: the voltage between the gate plane and the device channel at the position $x$ is the difference between the local voltage drop at $x$ ($V_b \cdot x/L$, $L$ is the channel length) and the voltage applied to the gate electrode $V_g$. The carrier density induced in the graphene by the field effect is then $n(x) = \left(\frac{V_b \cdot x}{L} - V_g\right) \cdot \frac{C_g}{e}$, where $C_g = \varepsilon\varepsilon_0/d_g$ is the gate capacitance, $\varepsilon$ and $\varepsilon_0$ are relative permittivity and electric constant, $d_g$ is the thickness of the gate dielectric, and $e$ is the elementary charge. The estimated charge carrier density induced by the self-gating effect is plotted as red dashed line in the figure 3b in the main text. It is interesting to note that self-gating effect sometimes strongly modifies current-voltage (*IV*) characteristics of graphene devices with large transconductance (large gate capacitance) leading to the kink or even negative differential resistance (NDR) region in *IV* curve [31]. The presence of NDR strongly depends on the gate capacitance, which in our case is 30 times smaller, thus preventing us from observing NDR [32, 33]. Furthermore, the channel length also affects the presence of NDR – the shorter is the channel the more pronounced is the NDR [34]. Ideally, the length-to-width ratio (L/W) should be below 0.2, while our Hall bar device has 10 times larger L/W, which again prevents us from observing the NDR.

Finally, thermally-induced carrier density scales with temperature as $T^4$, and at temperatures above 1000 K and at realistic gate voltages, the thermally-induced carriers dominate, figure S11. Since the electronic temperature forms a profile along the device channel (figure S8), the carrier density also varies correspondingly. This, together with intrinsic doping and bias induced carrier density (self-gating effect) may lead to the formation of p-n junctions and the asymmetry in current-voltage characteristics and in the temperature profile with the polarity of applied bias voltage (cf. positive and negative voltage branches in the figure 3, main text).

Interestingly, in the case of our transport measurements, the cryostat temperature was around 4 K while the temperature of electron liquid can rapidly reach ~700 K already at low electric fields. The current densities of 100-200 µA/µm corresponding to low electric fields on Figure 3c are sufficient to raise the temperature to 500-700 K. For example, using electrical currents of ~100 µA in a similar Hall bar devices the temperatures in excess of 300K were routinely obtained [35]. Further investigation of this interesting regime is beyond the scope of this article and comprises a separate study related to the hydrodynamic behaviour of electron fluid in graphene [36].

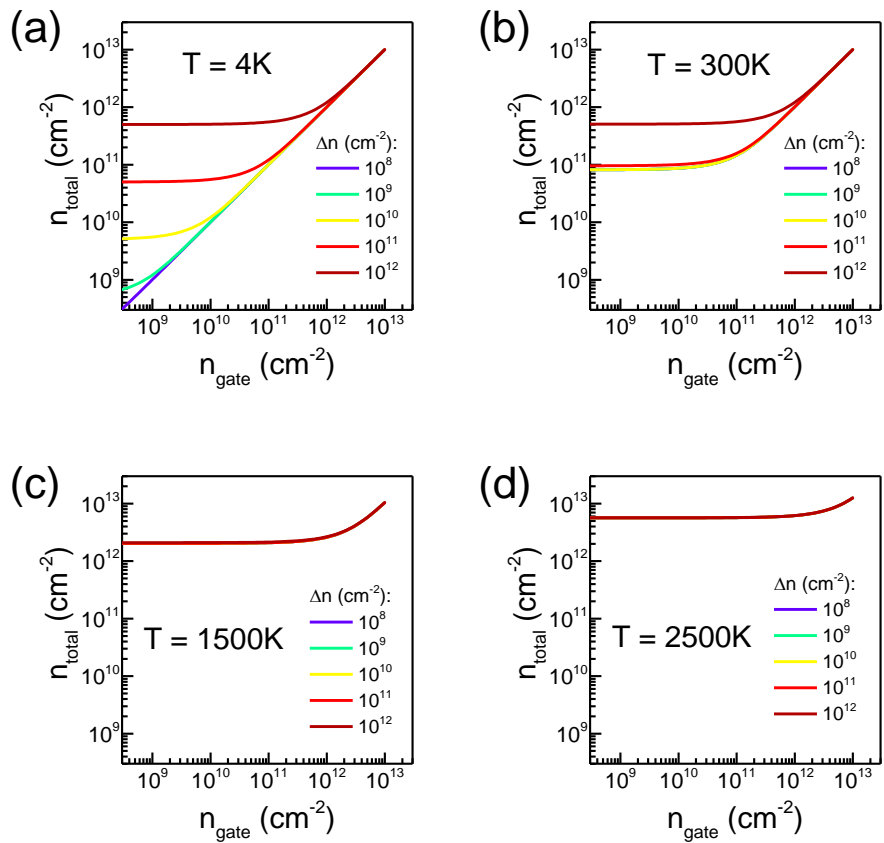

Figure S10. Total carrier density ($n_{total}$) vs gate-induced carrier density ($n_{gate}$) at different levels of disorder ($\Delta n$). Panels (a) to (d) show the effect of temperature smearing out the disorder-induced carrier density.

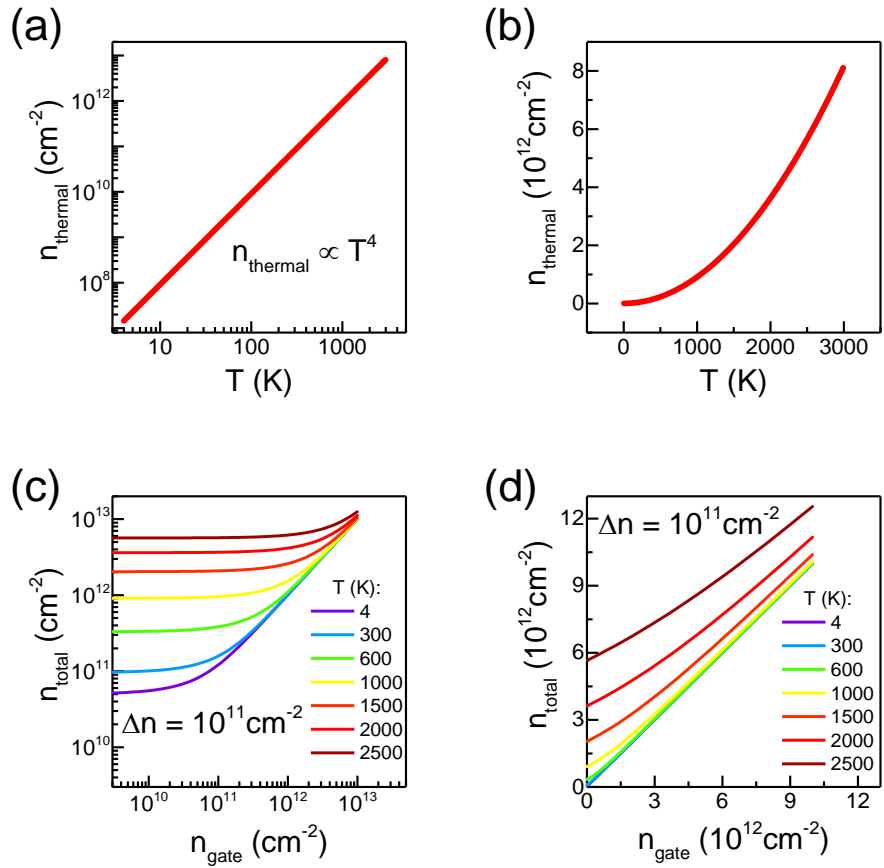

Figure S11. Total carrier density ($n_{total}$) vs gate-induced carrier density ($n_{gate}$) affected by thermally-induced carriers at different temperatures.